# QUBO formulations for numerical quantum computing

**Kyungtaek Jun**[a,*]

[a] Research Center, Innovative Quantum Computed Tomography, Seoul, Republic of Korea

[*] ktfriends@gmail.com

## Abstract

With the advent of quantum computers, many quantum computing algorithms are being developed. Solving linear systems is one of the most fundamental problems in almost all science and engineering. The Harrow-Hassidim-Lloyd algorithm, a monumental quantum algorithm for solving linear systems on gate model quantum computers, was invented and several advanced variations have been developed. For a given square matrix $A \in \mathbb{R}^{n \times n}$ and a vector $\vec{b} \in \mathbb{R}^n$, we will find unconstrained binary optimization (QUBO) models for a vector $\vec{x} \in \mathbb{R}^n$ that satisfies $A\vec{x} = \vec{b}$. To formulate QUBO models for a linear system solving problem, we make use of a linear least-square problem with binary representation of the solution. We validate those QUBO models on the D-Wave system and discuss the results. For a simple system, we provide a Python code to calculate the matrix characterizing the relationship between the variables and to print the test code that can be used directly in the D-Wave system.

## Introduction

Quantum computing has opened up a new paradigm for approaching computing problems where classical (i.e., traditional) computing can provide unparalleled speeds for specific problems. A specific subset of quantum computing is quantum annealing aimed at optimization problems. Quantum annealing processors naturally return low-energy solutions; some applications require the real minimum energy and others require good low-energy samples[1]. One popular model of optimization problems on which quantum annealers are based is the Ising model[2].

A monumental quantum algorithm for solving linear systems on gate model quantum computers, was invented in 2008, and a variety of advanced variations were developed[3]. At that time, their algorithm and its variants had many limitations, such as the quantum state of the solution, the need for quantum RAM, and the limitations of Matrix A[4]. In particular, the quantum state of the solution limits the use of the algorithm to submodules of the entire system, or reduces the quantum speed improvement in real world implementations, as it requires iterative computations to read the quantum state of each base. In 2019, a method of using linear least squares was proposed to solve a linear problem[5-7]. They suggested a QUBO model of linear equations for $m$ by $n$ linear systems. In addition, when solving a linear problem using the QUBO model, they compared several classic methods such as QR factorization and SVD[8]. There has been remarkable research[9-14] since the development of D-Wave quantum annealers.

In this paper, we propose two representative QUBO models for $n$ by $n$ linear systems. Additionally, we show the overall CPU calculations when performing the new QUBO modeling. The amount of calculation used here consists of simple addition and multiplication. These results will be of great advantage when solving linear equations for large $n$. We also provide Python code to create QUBO models that can be used in D-Wave quantum annealers. To date, the number of qubits that can be used in quantum annealers D-Wave 2000Q is approximately 2048, so the number of total qubit variable for the linear problem that can be solved using our QUBO model is 64. However, with the development of quantum annealers, the method we propose will have advantages. It is also expected that other issues could be approached in a similar way to ours.

# Method

## Background

The Ising model is a mathematical model for ferromagnetism in statistical mechanics. The energy Hamiltonian (the cost function) is formulated as follows:

$$\boldsymbol{H(\vec{\sigma}) = -\sum_{i=1}^{N} h_i\,\sigma_i - \sum_{i<j}^{N} J_{i,j}\,\sigma_i\sigma_j} \qquad \textbf{\textit{Equation 1}}$$

where $\vec{\sigma} = (\sigma_1, \cdots, \sigma_N)^T$ and $\sigma_i \in \{+1, -1\}$.

Quadratic unconstrained binary optimization (QUBO) is a combinatorial optimization problem in computer science. In this problem, a cost function $f$ is defined on an $n$-dimensional binary vector space $\mathbb{B}^n$ onto $\mathbb{R}$.

$$\boldsymbol{f(\vec{x}) = \vec{x}^T Q \vec{x}} \qquad \textbf{\textit{Equation 2}}$$

where $Q$ is an upper diagonal matrix, $\vec{x} = (x_1, \cdots, x_N)^T$, and $x_i \in \{0,1\}$. The problem is finding $x^*$, which minimizes the cost function $f$. Since we have $x_i^2 = x_i$, the cost function is reformulated as follows:

$$\boldsymbol{f(\vec{x}) = \sum_{i=1}^{N} Q_{i,i}\, x_i + \sum_{i<j}^{N} Q_{i,j}\, x_i x_j.} \qquad \textbf{\textit{Equation 3}}$$

In $Q$, the diagonal terms $Q_{i,i}$ and the off-diagonal terms $Q_{i,j}$ represent the linear terms and the quadratic terms, respectively. The unknowns of the Ising Model $\sigma$ and the unknowns of the QUBO Model $x$ have the linear relation

$$\boldsymbol{\sigma = 2x - 1} \qquad \textbf{\textit{Equation 4}}$$

Therefore, we can choose appropriate format depending on the binary unknown.

Given a matrix $A \in \mathbb{R}^{n \times n}$ and a column vector of variables $\vec{x} \in \mathbb{R}^n$ and a column vector $\vec{b} \in \mathbb{R}^n$. The linear least squares problem is to find the $\vec{x}$ that minimizes $\| A\vec{x} - \vec{b} \|$. In other words, it can be described as follows:

$$\arg\min_{\vec{x}} \| \boldsymbol{A\vec{x} - \vec{b}} \| \qquad \textbf{\textit{Equation 5}}$$

To solve Eq. 5 let us begin by writing out $\| A\vec{x} - \vec{b} \|$:

$$\boldsymbol{A\vec{x} - \vec{b} = \begin{pmatrix} a_{1,1} & a_{1,2} & \cdots & a_{1,n} \\ a_{2,1} & a_{2,2} & \cdots & a_{2,n} \\ \vdots & \vdots & \ddots & \vdots \\ a_{n,1} & a_{n,2} & \cdots & a_{n,n} \end{pmatrix} \begin{pmatrix} x_1 \\ x_2 \\ \vdots \\ x_n \end{pmatrix} - \begin{pmatrix} b_1 \\ b_2 \\ \vdots \\ b_n \end{pmatrix}} \qquad \textbf{\textit{Equation 6}}$$

Taking the 2-norm square of the resultant vector of Eq. 6, we get the following:

$$\| \boldsymbol{A\vec{x} - \vec{b}} \|_2^2 = \boldsymbol{\sum_{k=1}^{n} \left( \sum_{i=1}^{n} a_{k,i}\, x_i - b_k \right)^2} \qquad \textbf{\textit{Equation 7}}$$

$$\boldsymbol{= \sum_{k=1}^{n} \left( \left( \sum_{i=1}^{n} a_{k,i}\, x_i \right)^2 - 2b_k \sum_{i=1}^{n} a_{k,i}\, x_i + {b_k}^2 \right)} \qquad \textbf{\textit{Equation 8}}$$

$$\boldsymbol{= \sum_{k=1}^{n} \left( \sum_{i=1}^{n} \left(a_{k,i} x_i\right)^2 + 2 \sum_{i<j} a_{k,i}\, a_{k,j} x_i x_j - 2b_k \sum_{i=1}^{n} a_{k,i}\, x_i + {b_k}^2 \right)} \qquad \textbf{\textit{Equation 9}}$$

## QUBO modeling for linear system

If we were solving binary least squares, then each $x_i$ would be represented by the combination of qubits $q_{i,l} \in \{0,1\}$. In the work by O'Malley and Vesselinov[15], the radix 2 representation of positive real value $x_i$ is given by

$$\boldsymbol{x_i \approx \sum_{l=-m}^{m} 2^l\, q_{i,l}} \qquad \textit{Equation 10}$$

where positive integer $l$ is the number of digits of $x_i$ and negative $l$ is the number of digits of fractional terms.

Now, we can represent real value $x_i$ is below

$$\boldsymbol{x_i \approx \sum_{l=-m}^{m} 2^l\, q_{i,l}^{+} - \sum_{l=-m}^{m} 2^l\, q_{i,l}^{-}} \qquad \textit{Equation 11}$$

To represent both positive and negative numbers, $q_{i,l}^{+}$ and $q_{i,l}^{-}$ are involved. Note that this representation can have the same value with different binary combinations.

To drive QUBO model, we insert Eq. 11 into Eq. 9. We obtain the two summation terms of the first term in Eq. 9 as below:

$$\boldsymbol{\sum_{k=1}^{n}\sum_{i=1}^{n}\left(a_{k,i}x_i\right)^2 \approx \sum_{k=1}^{n}\sum_{i=1}^{n}\left(a_{k,i}^2\left(\sum_{l=-m}^{m} 2^l\, q_{i,l}^{+} - \sum_{l=-m}^{m} 2^l\, q_{i,l}^{-}\right)^2\right)} \qquad \textit{Equation 12}$$

$$\boldsymbol{= \sum_{k=1}^{n}\sum_{i=1}^{n}\left(a_{k,i}^2\left(\left(\sum_{l=-m}^{m} 2^l\, q_{i,l}^{+}\right)^2 + \left(\sum_{l=-m}^{m} 2^l\, q_{i,l}^{-}\right)^2\right)\right)} \qquad \textit{Equation 13}$$

$$\boldsymbol{= \sum_{k=1}^{n}\sum_{i=1}^{n} a_{k,i}^2\left(\sum_{l=-m}^{m} 2^{2l}\left(\left(q_{i,l}^{+}\right)^2 + \left(q_{i,l}^{-}\right)^2\right) + \sum_{l_1<l_2} 2^{l_1+l_2+1}\left(q_{i,l_1}^{+}q_{i,l_2}^{+} + q_{i,l_1}^{-}q_{i,l_2}^{-}\right)\right)} \qquad \textit{Equation 14}$$

$$\boldsymbol{= \sum_{k=1}^{n}\sum_{i=1}^{n}\sum_{l=-m}^{m} a_{k,i}^2\, 2^{2l}\left(q_{i,l}^{+} + q_{i,l}^{-}\right) + \sum_{k=1}^{n}\sum_{i=1}^{n}\sum_{l_1<l_2} a_{k,i}^2\, 2^{l_1+l_2+1}\left(q_{i,l_1}^{+}q_{i,l_2}^{+} + q_{i,l_1}^{-}q_{i,l_2}^{-}\right)} \qquad \textit{Equation 15}$$

Since $x_i$ is a positive, zero, or negative number, $q_{i,l_1}^{+} \times q_{i,l_2}^{-}$ for the same $i$ can be zero. We can get Eq. 13 from Eq. 12. Additionally, since $q^2 = q$, Eq. 15 can be finally obtained. Here, the front summation represents a linear term, and the second summation represents a quadratic term.

The QUBO form of the second term in Eq. 9 can be obtained as below:

$$\boldsymbol{\sum_{k=1}^{n}\sum_{i<j} 2\, a_{k,i}a_{k,j}x_ix_j \approx \sum_{k=1}^{n}\sum_{i<j} 2\, a_{k,i}a_{k,j}\left(\sum_{l=-m}^{m} 2^l\, q_{i,l}^{+} - \sum_{l=-m}^{m} 2^l\, q_{i,l}^{-}\right)\left(\sum_{l=-m}^{m} 2^l\, q_{j,l}^{+} - \sum_{l=-m}^{m} 2^l\, q_{j,l}^{-}\right)} \qquad \textit{Equation 16}$$

*Equation 17*

$$= \sum_{k=1}^{n} \sum_{i<j} \sum_{l_1=-m}^{m} \sum_{l_2=-m}^{m} 2^{l_1+l_2+1} a_{k,i} a_{k,j} (q_{i,l_1}^{+} q_{j,l_2}^{+} + q_{i,l_1}^{-} q_{j,l_2}^{-} - q_{i,l_1}^{+} q_{j,l_2}^{-} - q_{i,l_1}^{-} q_{j,l_2}^{+})$$

When we calculate the QUBO term, we use an upper triangular matrix. Therefore $q_{i,l}^{+}$ ,$q_{i,l}^{-}$,$q_{j,l}^{+}$, and $q_{j,l}^{-}$ have separate indices, and we can derive Eq. 17 from Eq. 16. This equation is part of the quadratic terms in our QUBO model.

The QUBO form of the third term in Eq. 9 can be obtained as below:

*Equation 18*

$$\sum_{k=1}^{n} \sum_{i=1}^{n} (-2 a_{k,i} b_k x_i) \approx \sum_{k=1}^{n} \sum_{i=1}^{n} \sum_{l=-m}^{m} 2^{l+1} a_{k,i} b_k (q_{i,l}^{-} - q_{i,l}^{+})$$

The above equation is part of the linear terms in our QUBO model.

The last term in Eq. 9 represents the minimum value when the vector $\vec{x}$ satisfies the linear equation in QUBO form. Therefore, our first QUBO model for linear systems is the summation of Eq. 15, Eq. 17, and Eq. 18.

To reduce the number of qubits to be used in Eq. 11, the new approximation of $x_i$ was introduced as follows:

*Equation 19*

$$x_i \approx -2^{m+1} q_i^{-} + \sum_{l=-m}^{m} 2^l q_{i,l}^{+}$$

Since the coefficient of $q_i^{-}$ represents the amount of translation of the range of the summation part in Eq. 19, any coefficient of $q_i^{-}$ that can represent the range of $x_i$ we want can be used. We select the coefficient of $q_i^{-}$, which represents the largest range that $x_i$ can have.

To drive QUBO model, we insert Eq. 19 into Eq. 9. We get another QUBO form of the first term in Eq. 9 as below:

*Equation 20*

$$\sum_{k=1}^{n} \sum_{i=1}^{n} (a_{k,i} x_i)^2 \approx \sum_{k=1}^{n} \sum_{i=1}^{n} \left( a_{k,i} \left( -2^{m+1} q_i^{-} + \sum_{l=-m}^{m} 2^l q_{i,l}^{+} \right) \right)^2$$

*Equation 21*

$$= \sum_{k=1}^{n} \sum_{i=1}^{n} a_{k,i}^2 \left( (-2^{m+1} q_i^{-})^2 + \sum_{l=-m}^{m} (-2^{l+m+2} q_i^{-} q_{i,l}^{+}) + \left( \sum_{l=-m}^{m} 2^l q_{i,l}^{+} \right)^2 \right)$$

*Equation 22*

$$= \sum_{k=1}^{n} \sum_{i=1}^{n} a_{k,i}^2 \left( 2^{2m+2} (q_i^{-})^2 + \sum_{l=-m}^{m} (-2^{l+m+2} q_i^{-} q_{i,l}^{+}) + \sum_{l=-m}^{m} 2^{2l} (q_{i,l}^{+})^2 + \sum_{l_1<l_2} 2^{l_1+l_2+1} q_{i,l_1}^{+} q_{i,l_2}^{+} \right)$$

*Equation 23*

$$= \sum_{k=1}^{n} \sum_{i=1}^{n} a_{k,i}^2 \left( 2^{2m+2} q_i^{-} + \sum_{l=-m}^{m} 2^{2l} q_{i,l}^{+} \right) + \sum_{k=1}^{n} \sum_{i=1}^{n} a_{k,i}^2 \left( \sum_{l=-m}^{m} (-2^{l+m+2} q_i^{-} q_{i,l}^{+}) + \sum_{l_1<l_2} 2^{l_1+l_2+1} q_{i,l_1}^{+} q_{i,l_2}^{+} \right)$$

Since $q^2 = q$, Eq. 23 can be finally obtained. The front summation represents linear terms, and the second summation represents quadratic terms.

The QUBO form of the second term in Eq. 9 can be obtained as below:

$$\sum_{k=1}^{n}\sum_{i<j} 2\, a_{k,i}a_{k,j}x_i x_j \approx \sum_{k=1}^{n}\sum_{i<j} 2\, a_{k,i}a_{k,j}\left(-2^{m+1}q_i^- + \sum_{l=-m}^{m} 2^l\, q_{i,l}^+\right)\left(-2^{m+1}q_j^- + \sum_{l=-m}^{m} 2^l\, q_{j,l}^+\right) \qquad \textit{Equation 24}$$

$$= \sum_{k=1}^{n}\sum_{i<j} a_{k,i}\, a_{k,j}\left(2^{2m+3}q_i^- q_j^- - \sum_{l=-m}^{m} 2^{l+m+2}\left(q_i^- q_{j,l}^+ + q_j^- q_{i,l}^+\right) + 2\sum_{l=-m}^{m} 2^l\, q_{i,l}^+ \sum_{l=-m}^{m} 2^l\, q_{j,l}^+\right) \qquad \textit{Equation 25}$$

$$= \sum_{k=1}^{n}\sum_{i<j} a_{k,i}\, a_{k,j}\left(2^{2m+3}q_i^- q_j^- - \sum_{l=-m}^{m} 2^{l+m+2}\left(q_i^- q_{j,l}^+ + q_j^- q_{i,l}^+\right) + \sum_{l_1=-m}^{m}\sum_{l_2=-m}^{m} 2^{l_1+l_2+1}\, q_{i,l_1}^+ q_{j,l_2}^+\right) \qquad \textit{Equation 26}$$

Equation 26 can be derived in a similar way to Eq. 17. This equation is part of the quadratic terms in our second QUBO model.

The QUBO form of the third term in Eq. 9 can be obtained as below:

$$\sum_{k=1}^{n}\sum_{i=1}^{n}\left(-2a_{k,i}b_k x_i\right) \approx \sum_{k=1}^{n}\sum_{i=1}^{n} 2^{m+2}\, a_{k,i}b_k q_{i,l}^+ + \sum_{k=1}^{n}\sum_{i=1}^{n}\sum_{l=-m}^{m}\left(-2^{l+1}a_{k,i}b_k q_{i,l}^+\right) \qquad \textit{Equation 27}$$

The last term in Eq. 9 represents the minimum value when the vector $\vec{x}$ satisfies the linear equation in QUBO form.

Therefore, our second QUBO model for linear systems is the summation of Eq. 23, Eq. 26, and Eq. 27.

## QUBO modeling for eigenvalues and eigenvectors

Given a matrix $A \in \mathbb{R}^{n\times n}$ and a column vector of variables $\vec{x} \in \mathbb{R}^n$ and a real number $\boldsymbol{\lambda} \in \mathbb{R}$. The linear least squares problem is to find the $\boldsymbol{\lambda}$ and $\vec{x}$ that satisfies $A\vec{x} = \boldsymbol{\lambda}\vec{\boldsymbol{x}}$. In other words, it can be described as follows:

$$\arg\min_{\boldsymbol{\lambda},\vec{x}} \| \boldsymbol{A}\vec{\boldsymbol{x}} - \boldsymbol{\lambda}\vec{\boldsymbol{x}} \| = \boldsymbol{0} \qquad \textit{Equation 28}$$

To solve Eq. 28 let us begin by writing out $\| A\vec{x} - \boldsymbol{\lambda}\vec{\boldsymbol{x}} \|$:

$$\boldsymbol{A}\vec{\boldsymbol{x}} - \boldsymbol{\lambda}\vec{\boldsymbol{x}} = \begin{pmatrix} a_{1,1} & a_{1,2} & \cdots & a_{1,n} \\ a_{2,1} & a_{2,2} & \cdots & a_{2,n} \\ \vdots & \vdots & \ddots & \vdots \\ a_{n,1} & a_{n,2} & \cdots & a_{n,n} \end{pmatrix}\begin{pmatrix} x_1 \\ x_2 \\ \vdots \\ x_n \end{pmatrix} - \boldsymbol{\lambda}\begin{pmatrix} x_1 \\ x_2 \\ \vdots \\ x_n \end{pmatrix} \qquad \textit{Equation 29}$$

Taking the 2-norm square of the resultant vector of Eq. 10, we get the following:

$$\| \boldsymbol{A}\vec{\boldsymbol{x}} - \boldsymbol{\lambda}\vec{\boldsymbol{x}} \|_2^2 = \vec{\boldsymbol{x}}^T\boldsymbol{A}^T\boldsymbol{A}\vec{\boldsymbol{x}} - 2\boldsymbol{\lambda}\vec{\boldsymbol{x}}^T\boldsymbol{A}\vec{\boldsymbol{x}} + \boldsymbol{\lambda}^2\vec{\boldsymbol{x}}^T\vec{\boldsymbol{x}} \qquad \textit{Equation 30}$$

If we were solving binary least squares, then each $x_i$ would be represented by the combination of qubits $q_{i,l} \in \{0,1\}$. In the work by O'Malley and Vesselinov[11], the radix 2 representation of positive real value $\boldsymbol{\lambda}$ is given by

$$\lambda \approx \sum_{l=-m}^{m} 2^l q_l^+ - \sum_{l=-m}^{m} 2^l q_l^- \qquad \textit{Equation 31}$$

where positive integer $l$ is the number of digits of $x_i$ and negative $l$ is the number of digits of fractional terms.

To drive QUBO model, we insert Eq. 31 into Eq. 30. We obtain the two summation terms of the first term in Eq. 30 as below:

$$\vec{x}^T A^T A \vec{x} = \sum_{k=1}^{n} \left( \sum_{i=1}^{n} (a_{k,i} x_i)^2 + 2 \sum_{i<j} a_{k,i}\, a_{k,j} x_i x_j \right) \qquad \textit{Equation 32}$$

**Eq. 32 can be calculated as follows:**

$$\sum_{k=1}^{n} \sum_{i=1}^{n} (a_{k,i} x_i)^2 \approx \sum_{k=1}^{n} \sum_{i=1}^{n} \left( a_{k,i}^2 \left( \sum_{l=0}^{m} 2^l q_{i,l}^+ - \sum_{l=0}^{m} 2^l q_{i,l}^- \right)^2 \right) \qquad \textit{Equation 33}$$

$$= \sum_{k=1}^{n} \sum_{i=1}^{n} \left( a_{k,i}^2 \left( \left( \sum_{l=0}^{m} 2^l q_{i,l}^+ \right)^2 + \left( \sum_{l=0}^{m} 2^l q_{i,l}^- \right)^2 \right) \right) \qquad \textit{Equation 34}$$

$$= \sum_{k=1}^{n} \sum_{i=1}^{n} a_{k,i}^2 \left( \sum_{l=0}^{m} 2^{2l} \left( (q_{i,l}^+)^2 + (q_{i,l}^-)^2 \right) + \sum_{l_1<l_2} 2^{l_1+l_2+1} \left( q_{i,l_1}^+ q_{i,l_2}^+ + q_{i,l_1}^- q_{i,l_2}^- \right) \right) \qquad \textit{Equation 35}$$

$$= \sum_{k=1}^{n} \sum_{i=1}^{n} \sum_{l=0}^{m} a_{k,i}^2\, 2^{2l} \left( q_{i,l}^+ + q_{i,l}^- \right) + \sum_{k=1}^{n} \sum_{i=1}^{n} \sum_{l_1<l_2} a_{k,i}^2\, 2^{l_1+l_2+1} \left( q_{i,l_1}^+ q_{i,l_2}^+ + q_{i,l_1}^- q_{i,l_2}^- \right) \qquad \textit{Equation 36}$$

$$\sum_{k=1}^{n} \sum_{i<j} 2\, a_{k,i} a_{k,j} x_i x_j \approx \sum_{k=1}^{n} \sum_{i<j} 2\, a_{k,i} a_{k,j} \left( \sum_{l=0}^{m} 2^l q_{i,l}^+ - \sum_{l=0}^{m} 2^l q_{i,l}^- \right) \left( \sum_{l=0}^{m} 2^l q_{j,l}^+ - \sum_{l=0}^{m} 2^l q_{j,l}^- \right) \qquad \textit{Equation 37}$$

$$= \sum_{k=1}^{n} \sum_{i<j} \sum_{l_1=0}^{m} \sum_{l_2=0}^{m} 2^{l_1+l_2+1} a_{k,i} a_{k,j} \left( q_{i,l_1}^+ q_{j,l_2}^+ + q_{i,l_1}^- q_{j,l_2}^- - q_{i,l_1}^+ q_{j,l_2}^- - q_{i,l_1}^- q_{j,l_2}^+ \right) \qquad \textit{Equation 38}$$

Since $x_i$ is a positive, zero, or negative number, $q_{i,l_1}^+ \times q_{i,l_2}^-$ for the same $i$ can be zero. We can get Eq. 34 from Eq. 33. Additionally, since $q^2 = q$, Eq. 36 can be finally obtained. Here, the front summation represents a linear term, and the second summation represents a quadratic term. Therefore $q_{i,l}^+$ ,$q_{i,l}^-$,$q_{j,l}^+$, and $q_{j,l}^-$ have separate indices, and we can derive Eq. 38 from Eq. 37. This equation is part of the quadratic terms in our QUBO model.

The QUBO form of the second term in Eq. 30 can be obtained as below:

$$-2\lambda \vec{x}^T A \vec{x} \approx -2 \left( \sum_{l=-m}^{m} 2^l q_l^+ - \sum_{l=-m}^{m} 2^l q_l^- \right) \sum_{k_2=1}^{n} \sum_{k_1=1}^{n} \left( a_{k_2,k_1} \left( \sum_{l=0}^{m} 2^l q_{k_1,l}^+ - \sum_{l=0}^{m} 2^l q_{k_1,l}^- \right) \left( \sum_{l=0}^{m} 2^l q_{k_2,l}^+ - \sum_{l=0}^{m} 2^l q_{k_2,l}^- \right) \right) \qquad \textit{Equation 39}$$

*Equation 40*

$$= \left(\sum_{l=-m}^{m} 2^{l+1} (q_l^+ - q_l^-)\right) \sum_{k_2=1}^{n} \sum_{k_1=1}^{n} \left( a_{k_2,k_1} \left( \left(\sum_{l=0}^{m} 2^l q_{k_1,l}^+\right)\left(\sum_{l=0}^{m} 2^l q_{k_2,l}^-\right) + \left(\sum_{l=0}^{m} 2^l q_{k_1,l}^-\right)\left(\sum_{l=0}^{m} 2^l q_{k_2,l}^+\right) - \left(\sum_{l=0}^{m} 2^l q_{k_1,l}^+\right)\left(\sum_{l=0}^{m} 2^l q_{k_2,l}^+\right) - \left(\sum_{l=0}^{m} 2^l q_{k_1,l}^-\right)\left(\sum_{l=0}^{m} 2^l q_{k_2,l}^-\right) \right)\right)$$

*Equation 41*

$$= \sum_{k_1=1}^{n} \sum_{k_2=1}^{n} \sum_{l=-m}^{m} \sum_{l_1=0}^{m} \sum_{l_2=0}^{m} \left( 2^{l+l_1+l_2+1} a_{k_2,k_1} ( q_l^+ q_{k_1,l_1}^+ q_{k_2,l_2}^- + q_l^+ q_{k_1,l_1}^- q_{k_2,l_2}^+ - q_l^+ q_{k_1,l_1}^+ q_{k_2,l_2}^+ - q_l^+ q_{k_1,l_1}^- q_{k_2,l_2}^- - q_l^- q_{k_1,l_1}^+ q_{k_2,l_2}^- - q_l^- q_{k_1,l_1}^- q_{k_2,l_2}^+ + q_l^- q_{k_1,l_1}^+ q_{k_2,l_2}^+ + q_l^- q_{k_1,l_1}^- q_{k_2,l_2}^- )\right)$$

This equation is part of the cubic terms in our QUBO model.

The QUBO form of the third term in Eq. 30 can be obtained as below:

$$\lambda^2 \vec{x}^T \vec{x} \approx \left(\sum_{l=-m}^{m} 2^l q_l^+ - \sum_{l=-m}^{m} 2^l q_l^-\right)^2 \left( \sum_{k=1}^{n} \left( \sum_{l=0}^{m} 2^l q_{k,l}^+ - \sum_{l=0}^{m} 2^l q_{k,l}^- \right)^2 \right)$$

*Equation 42*

*Equation 43*

$$= \left( \sum_{l=-m}^{m} 2^{2l} ((q_l^+)^2 + (q_l^-)^2) + \sum_{l_1<l_2} 2^{l_1+l_2+1} (q_{l_1}^+ q_{l_2}^+ + q_{l_1}^- q_{l_2}^-) \right) \left( \sum_{k=1}^{n} \left( \sum_{l=0}^{m} 2^{2l} \left( (q_{k,l}^+)^2 + (q_{k,l}^-)^2 \right) + \sum_{l_1<l_2} 2^{l_1+l_2+1} (q_{k,l_1}^+ q_{k,l_2}^+ + q_{k,l_1}^- q_{k,l_2}^-) \right)\right)$$

*Equation 44*

$$= \sum_{k_1=1}^{n} \left( \sum_{l_1=-m}^{m} \sum_{l_2=0}^{m} 2^{2(l_1+l_2)} ( q_{l_1}^+ q_{k,l_2}^+ + q_{l_1}^+ q_{k,l_2}^- + q_{l_1}^- q_{k,l_2}^+ + q_{l_1}^- q_{k,l_2}^- ) + \sum_{l_1=-m}^{m} \sum_{0\le l_2<l_3\le m} 2^{2l_1+l_2+l_3+1} ( q_{l_1}^+ q_{k,l_2}^+ q_{k,l_3}^+ + q_{l_1}^+ q_{k,l_2}^- q_{k,l_3}^- + q_{l_1}^- q_{k,l_2}^+ q_{k,l_3}^+ + q_{l_1}^- q_{k,l_2}^- q_{k,l_3}^- ) + \sum_{l_1=0}^{m} \sum_{-m\le l_2<l_3\le m} 2^{2l_1+l_2+l_3+1} ( q_{k,l_1}^+ q_{l_2}^+ q_{l_3}^+ + q_{k,l_1}^+ q_{l_2}^- q_{l_3}^- + q_{k,l_1}^- q_{l_2}^+ q_{l_3}^+ + q_{k,l_1}^- q_{l_2}^- q_{l_3}^- ) + \sum_{-m\le l_1<l_2\le m} \sum_{0\le l_3<l_4\le m} 2^{l_1+l_2+l_3+l_4+2} ( q_{l_1}^+ q_{l_2}^+ q_{k,l_3}^+ q_{k,l_4}^+ + q_{l_1}^+ q_{l_2}^+ q_{k,l_3}^- q_{k,l_4}^- + q_{l_1}^- q_{l_2}^- q_{k,l_3}^+ q_{k,l_4}^+ + q_{l_1}^- q_{l_2}^- q_{k,l_3}^- q_{k,l_4}^- ) \right)$$

The above equation consists of quadratic, cubic, and quartic terms.

To reformulate a non-quadratic (higher-degree) polynomial to Ising/QUBO, substitute terms in the form of $axyz$, where $a$ is a real number, with one of the following quadratic terms[16]:

$$\boldsymbol{axyz} = \begin{cases} \boldsymbol{aw(x+y+z-2), a<0} \\ \boldsymbol{a\{w(x+y+z-1)+(xy+yz+zx)-(x+y+z)+1\}}, \quad \boldsymbol{a>0} \end{cases}$$ ***Equation 45***

Equation 28 satisfies the minimum value 0 when the variables, scholar $\boldsymbol{\lambda}$ and vector $\vec{x}$, satisfy the eigenvalue and eigenvector in QUBO form. Therefore, our first QUBO model for eigenvalues and eigenvectors will be obtained by polynomial reduction by minimum selection in Eq. 45 after the summation of Eq. 36, Eq. 38, Eq. 41, and Eq. 44.

To reduce the number of qubits to be used in Eq. 11, the new approximation of $x_i$ was introduced as follows:

$$\boldsymbol{x_i \approx -2^{m+1} q_i^- + \sum_{l=-m}^{m} 2^l\, q_{i,l}^+}$$ ***Equation 46***

The QUBO model can be derived in the same way as above.

# Implementation for linear system

Let's consider the following example for linear system[17]:

$$\boldsymbol{A} = \begin{pmatrix} \boldsymbol{3} & \boldsymbol{1} \\ \boldsymbol{-1} & \boldsymbol{2} \end{pmatrix}, \ \vec{\boldsymbol{b}} = \begin{pmatrix} \boldsymbol{-1} \\ \boldsymbol{5} \end{pmatrix}, \ \text{and} \ \vec{\boldsymbol{x}} = \begin{pmatrix} \boldsymbol{x_1} \\ \boldsymbol{x_2} \end{pmatrix} = \begin{pmatrix} \boldsymbol{q_{11}+2q_{12}-q_{13}-2q_{14}} \\ \boldsymbol{q_{21}+2q_{22}-q_{23}-2q_{24}} \end{pmatrix}$$ ***Equation 47***

where $q_{ij}$ is binary and the range of integer $x_i$ is $-3 \le x_i \le 3$.

They used the related matrix below to solve Eq. 28.

$$\boldsymbol{Q_1} = \begin{pmatrix} 26 & 40 & -20 & -40 & 2 & 4 & -2 & -4 \\ & 72 & -40 & -80 & 4 & 8 & -4 & -8 \\ & & -6 & 40 & -2 & -4 & 2 & 4 \\ & & & 8 & -4 & -8 & 4 & 8 \\ & & & & -13 & 20 & -10 & -20 \\ & & & & & -16 & -20 & -40 \\ & & & & & & 23 & 20 \\ & & & & & & & 56 \end{pmatrix}$$ ***Equation 48***

The diagonal entries and the upper diagonal entries represent coefficients of linear terms and quadratic terms, respectively, in Formula[2].

Our matrix from QUBO modeling 1 is below:

$$\boldsymbol{Q_2} = \begin{pmatrix} 26 & 40 & 0 & 0 & 2 & 4 & -2 & -4 \\ & 72 & 0 & 0 & 4 & 8 & -4 & -8 \\ & & -6 & 40 & -2 & -4 & 2 & 4 \\ & & & 8 & -4 & -8 & 4 & 8 \\ & & & & -13 & 20 & 0 & 0 \\ & & & & & -16 & 0 & 0 \\ & & & & & & 23 & 20 \\ & & & & & & & 56 \end{pmatrix} \quad \textit{Equation 49}$$

Two different matrices have the same solution for Eq. 28.

## Empirical Test

We tested the QUBO model of Eq. 30 on the D-Wave 2000Q system with $10{,}000$ anneals, and compared the result from Eq. 29[16]. List 1 shows the Python code for our matrix and prints the Python code for running the QUBO model described in Formula 30. This code creates a QUBO model with nonzero coefficient terms.

```
import numpy as np
import random, math
import copy

Dimension = 2
qubits = 2
A = np.array([[3, 1], [-1, 2]])
b = np.array([-1, 5])

QM = np.zeros((2*qubits*Dimension, 2*qubits*Dimension))
for k in range(Dimension):
    for i in range(Dimension):
        for l in range(qubits):
            cef1 = pow(2,2*l)*pow(A[k][i],2)
            cef2 = pow(2,l+1)*A[k][i]*b[k]
            po1 = 4*i + l
            po2 = 4*i + l + 2
            QM[po1][po1] = QM[po1][po1] + cef1 - cef2
            QM[po2][po2] = QM[po2][po2] + cef1 + cef2

for k in range(Dimension):
    for i in range(Dimension):
        for l1 in range(qubits-1):
            for l2 in range(l1+1,qubits):
                qcef = pow(2, l1+l2+1)*pow(A[k][i],2)
                po1 = 4*i + l1
                po2 = 4*i + l2
                QM[po1][po2] = QM[po1][po2] + qcef
                po3 = 4*i + l1 + 2
                po4 = 4*i + l2 + 2
                QM[po3][po4] = QM[po3][po4] + qcef

for k in range(Dimension):
    for i in range(Dimension-1):
        for j in range(i+1,Dimension):
            for l1 in range(qubits):
                for l2 in range(qubits):
                    qcef = pow(2, l1+l2+1)*A[k][i]*A[k][j]
                    po1 = 4*i + l1
                    po2 = 4*j + l2
                    QM[po1][po2] = QM[po1][po2] + qcef
                    po3 = 4*i + l1 + 2
                    po4 = 4*j + l2 + 2
                    QM[po3][po4] = QM[po3][po4] + qcef
                    po5 = 4*i + l1
                    po6 = 4*j + l2 + 2
                    QM[po5][po6] = QM[po5][po6] - qcef
                    po7 = 4*i + l1 + 2
                    po8 = 4*j + l2
                    QM[po7][po8] = QM[po7][po8] - qcef

# Print Matrix Q
print("# Matrix Q is")
print(QM)
```

```
# Print Python code for the run in D-Wave quantum processing unit
print("from dwave.system import DWaveSampler, EmbeddingComposite")
print("sampler_auto = EmbeddingComposite(DWaveSampler(solver={'qpu': True}))\n")
print("linear = {", end = "")
for i in range(2*qubits*Dimension-1):
    linear = i + 1
    print ("('q",linear,"','q",linear,"'):",format(QM[i][i]),sep='', end = ', ')
print ("('q",2*qubits*Dimension,"','q",2*qubits*Dimension,"'):",format(QM[2*qubits*Dimension-1][2*qubits*Dimension-1]),"}", sep='')

print("\nquadratic = {", end = "")
for i in range(2*qubits*Dimension-1):
    for j in range(i+1,2*qubits*Dimension):
        if QM[i][j] != 0:
            qdrt1 = i + 1
            qdrt2 = j + 1
            if i == 2*qubits*Dimension-2 and j == 2*qubits*Dimension-1:
                print ("('q",qdrt1,"','q",qdrt2,"'):",format(QM[i][j]), "}", sep='')
            else:
                print ("('q",qdrt1,"','q",qdrt2,"'):",format(QM[i][j]), sep ='', end = ', ')

print("\nQ = dict(linear)")
print("Q.update(quadratic)\n")

qa_iter = 1000
print("sampleset = sampler_auto.sample_qubo(Q, num_reads=",qa_iter,")", sep = "")
print("print(sampleset)")
```

**List 1**. Python code for 2 × 2 Matrix Q and the run in the D-Wave quantum processing unit (QPU)

For $Q_1$, Table 1 shows the number of occurrences of the lowest energy $-26$ for invertible matrix $A$. The total occurrence of all combinations of qubits having the lowest energy is $6{,}316$ from $10{,}000$ anneals. In Eq. 9, we have the constant term $26$ and the QUBO form in Eq. 29 is the same with Eq. 9 except the constant term. Since the possible minimum of Eq. 9 is zero, the possible minimum of our QUBO model is $-26$ calculated as the summation of $b_k$. For $x_1 = -1$, there are three qubit representations: $[q_{11}, q_{12}, q_{13}, q_{14}] =$ [0,0,1,0], [0,1,1,1] or [1,0,0,1]. For $x_2 = 2$, there are two qubit representations: $[q_{21}, q_{22}, q_{23}, q_{24}] =$ [0,1,0,0] or [1,1,1,0]. Thus, 6,316 anneals achieved the lowest energy state, with an energy equal to -26.

| $q_{11}$ | $q_{12}$ | $q_{13}$ | $q_{14}$ | $q_{21}$ | $q_{22}$ | $q_{23}$ | $q_{24}$ | energy | # of occurrences |
|---|---|---|---|---|---|---|---|---|---|
| 0 | 1 | 1 | 1 | 0 | 1 | 0 | 0 | -26.0 | 959 |
| 0 | 1 | 1 | 1 | 1 | 1 | 1 | 0 | -26.0 | 2066 |
| 0 | 0 | 1 | 0 | 0 | 1 | 0 | 0 | -26.0 | 499 |
| 1 | 0 | 0 | 1 | 1 | 1 | 1 | 0 | -26.0 | 1256 |
| 0 | 0 | 1 | 0 | 1 | 1 | 1 | 0 | -26.0 | 925 |
| 1 | 0 | 0 | 1 | 0 | 1 | 0 | 0 | -26.0 | 611 |

**Table 1.** The Results of the solutions for the QUBO model, Eq. 29

For $Q_2$ with zero terms in QUBO, Table 2 shows the number of occurrences of the lowest energy $-26$. The total occurrence of all combinations of qubits having the lowest energy is $8{,}400$ from $10{,}000$ anneals. Thus, 8,400 anneals achieved the lowest energy state, with an energy equal to -26.

| $q_{11}$ | $q_{12}$ | $q_{13}$ | $q_{14}$ | $q_{21}$ | $q_{22}$ | $q_{23}$ | $q_{24}$ | energy | # of occurrences |
|---|---|---|---|---|---|---|---|---|---|
| 0 | 0 | 1 | 0 | 0 | 1 | 0 | 0 | -26.0 | 8400 |

**Table 2.** The Results of the solutions for the QUBO model, Eq. 29 with zero terms

For $Q2$ without zero terms in QUBO, Table 3 shows the number of occurrences of the lowest energy $-26$. The total occurrence of all combinations of qubits having the lowest energy is $8{,}725$ from $10{,}000$ anneals. Thus, 8,725 anneals achieved the lowest energy state, with an energy equal to -26.

| $q_{11}$ | $q_{12}$ | $q_{13}$ | $q_{14}$ | $q_{21}$ | $q_{22}$ | $q_{23}$ | $q_{24}$ | energy | # of occurrences |
|---|---|---|---|---|---|---|---|---|---|
| 0 | 0 | 1 | 0 | 0 | 1 | 0 | 0 | -26.0 | 8725 |

**Table 3.** The Results of the solutions for the QUBO model, Eq. 29 without zero terms

# Discussion

The most computationally intensive part of our first QUBO form is Eq. 17, one of the quadratic terms. We want to discuss Eq. 17 below. By computing Eq. 17, we calculate variables $k$,$i$, and $j$. Variable $k$ can vary from 1 to $n$, and $i$ and $j$ represent the $(i, j)$ components of the upper triangular matrices without diagonal components, so the total amount of calculation can be expressed as follows:

***Equation 50***

Major computing cost for $(\boldsymbol{i}, \boldsymbol{j}) = \frac{\boldsymbol{n(n-1)}}{\boldsymbol{2}}$

$l_1 + l_2$ can have a value from $-2m$ to $2m$. Additionally, the product of $a_{k,i}a_{k,j}$ and $2^{l_1+l_2+1}$ can be calculated by addition. Therefore, the total amount of calculation is as follows:

***Equation 51***

Major computing cost for $(\boldsymbol{i}, \boldsymbol{j}, \boldsymbol{l_1}, \boldsymbol{l_2}) = (\boldsymbol{4m+1}) + \frac{\boldsymbol{n(n-1)}}{\boldsymbol{2}}$

***Equation 52***

Major computing cost for $(\boldsymbol{i}, \boldsymbol{j}, \boldsymbol{k}, \boldsymbol{l_1}, \boldsymbol{l_2}) = (\boldsymbol{4m+1})\boldsymbol{n} + \frac{\boldsymbol{n^2(n-1)}}{\boldsymbol{2}}$

To make the QUBO model solve the linear equation, the total amount of calculation seems to be large as shown in Eq. 32. However, since each step consists of one multiplication or one addition, the solution can be solved faster than solving a linear equation in a classical computer. Our Python code List 1 did not minimize the amount of computation to improve readability.

To remove the fractional part of $x_i$ in Eqs. 11 and 19, we can use $\| A(c\vec{x}) - c\vec{b} \|$instead of $\| A\vec{x} - \vec{b} \|$. In more detail, we can use the $\| A\vec{y} - c\vec{b} \|$ equation and $y_i \approx \sum_{l=0}^{m} 2^l\, q_{i,l}$ for equation $\| A(c\vec{x}) - c\vec{b} \|$. If the solution requires an accuracy of one hundredth in the fractional part, then c is 100. In Eq. 9, $b_k$ can be changed to $cb_k$. In the case of using the QUBO form to find a linear solution in quantum annealing, we already know the minimum value

for the invertible matrix. For example, if the range of x in Eq. 28 is expressed as a number with two decimal places and two digits, then the positive value of $x_i$ must be expressed as $x_i \approx \sum_{l=-6}^{6} 2^l\, q_{i,l}^+$. For the new formula $\| A\vec{y} - c\vec{b} \|$, the positive value of $y_i$ must be expressed as $y_i = \sum_{l=0}^{13} 2^l\, q_{i,l}^+$. In this case, one more qubit is used, but a more accurate solution can be obtained.

The greatest advantage of this model lies in parallel computing in the process of creating the QUBO equation. We can transform Eq. 17 as follows to calculate the cost for parallel computing:

$$\sum_{l_1=-m}^{m} \sum_{l_2=-m}^{m} \left( \sum_{k=1}^{n} \sum_{i<j} 2^{l_1+l_2+1}\, a_{k,i} a_{k,j} \right) \left( q_{i,l_1}^+ q_{j,l_2}^+ + q_{i,l_1}^- q_{j,l_2}^- - q_{i,l_1}^+ q_{j,l_2}^- - q_{i,l_1}^- q_{j,l_2}^+ \right) \quad \textit{Equation 53}$$

In Eq. 34, the coefficient of each quadratic term is divisible with respect to $k$. This means that it can be efficiently partitioned when distributing n parallel processes. At this time, each process requires computing cost as in Eq. 32. To calculate the remaining term the fastest, it can be calculated through $n(n-1)/4$ parallel systems for each $k$. In this case, the efficiency of the total computational cost is reduced, but the computational cost of the QUBO model is approximately $2log_2 n$. When performing our QUBO modeling, as the total number of processes increases, the computing cost decreases. However, there is a large drawback to calculating linear equations in a quantum annealer. Our QUBO formula requires that one qubit be connected to all other qubits. However, each quantum annealer has its own topology, which limits the connectivity between qubits. To develop a QUBO model that can be used in a quantum annealer where the connection between qubits is limited, it is necessary to study the method of diagonalizing $Q_2$ based on a quantum annealer's topology.

QUBO modeling 1 has a major disadvantage in that it uses approximately twice as many qubits as QUBO modeling 2. However, we mainly discussed the results of QUBO modeling 1 to explain the important results for the constraints of the QUBO models. To solve the system of linear equations, each element $x_i$ of $\vec{x}$ can be divided by positive, negative, or zero. When Eq. 11 is used for the element $x_i$, and the result can be expressed as combinations of elements in $q^+$ and $q^-$. However, a solution $x_i$ can be made with either $\sum 2^l q_{i,l}^+$ or $-\sum 2^l q_{i,l}^-$. Because of this property, we can assume this problem to be a constrained problem for certain quadratic terms. To make certain quadratic terms 0, it is common to add the coefficients of the terms to a large value. To make certain quadratic terms equal to 0, it is common to add the coefficients of the terms to a large value. However, in this paper, we propose a new method of adding different values to each term so that it is set to 0 when the coefficient of each term is negative. Therefore, we set each coefficient of $q_{i,l_1}^{+} q_{i,l_2}^{-}$ to zero for each $x_i$. Equation 30 was made by adding the same positive value to the coefficient of each negative constrained term in Eq. 29. In this way, we experimented by adding values of 100, 500, and 1000 to the constraint terms of Q matrices of various sizes with a size of 64 by 64 or less. However, we did not obtain better results than the two methods of setting each constrained term to 0 introduced in this paper. We obtained these results in D-wave 2000Q. The reason why we obtained the best results by removing the constrained terms from QUBO is probably because we obtained a physical advantage by simplifying the superposition and entanglement of qubits by reducing the terms.

In the cubic and quartic terms of Eqs. 41 and 4, polynomial reduction occurs in different forms depending on the sign of the coefficient[17]. Since the eigenvalue is positive, negative, or 0, we can assume that the eigenvalue is positive or negative in Eq. 31 and create two QUBO models. When an eigenvalue and an eigenvector having a minimum value of 0 in each QUBO model are expressed, they become the eigenvalues and eigenvectors of the actual matrix.

*Additional results: "https://github.com/ktfriends/Quantum_Computing/tree/main/QUBO_Ising"*

# Author contributions Statement

K. J conceived and designed the research problem and coded for simulations using Python Numpy. K.J performed the hardware testing. All authors developed theoretical results and wrote the paper.

# Competing financial interests

The authors declare no competing financial and non-financial interests.